\documentclass[journal=jpccck,manuscript=article,layout=traditional]{achemso}

\usepackage{graphicx}
\usepackage[intlimits]{amsmath}
\usepackage{amssymb}
\usepackage{amsthm}
\usepackage{booktabs}
\usepackage{multirow}
\usepackage{natbib}
\usepackage{geometry}
\usepackage{color}
\usepackage{bm}

\author{R. Carmina Monreal}
\affiliation{Departamento de F\'{\i}sica Te\'orica de la Materia Condensada C5 and Condensed Matter Physics Center (IFIMAC), Universidad Aut\'onoma de Madrid, E-28049 Madrid, Spain}
\email{r.c.monreal@uam.es}
\author{Tomasz J. Antosiewicz}
\affiliation{Centre of New Technologies, University of Warsaw, Banacha 2c, 02-097 Warsaw, Poland}
\alsoaffiliation{Department of Physics and Gothenburg Physics Centre, Chalmers University of Technology, SE-412 96 G\"oteborg, Sweden}
\author{S. Peter Apell}
\affiliation{Department of Physics and Gothenburg Physics Centre, Chalmers University of Technology, SE-412 96 G\"oteborg, Sweden}

\title{Diffuse Surface Scattering and Quantum Size Effects in the Surface Plasmon Resonances of Low Carrier Density Nanocrystals}


\begin{document}

\begin{abstract}
The detailed understanding of the physical parameters that determine Localized Surface Plasmon Resonances (LSPRs) is essential to develop new applications for plasmonics. A relatively new area of research has been opened by the identification of LSPRs in low carrier density systems obtained by doping semiconductor quantum dots. 
We investigate theoretically how diffuse surface scattering of electrons in combination with the effect of quantization due to size (QSE) impact 
the evolution of the LSPRs with the size of these nanosystems.
Two key parameters are the length $R_0$ giving the strength of the QSE and the velocity $\beta_T$ of the electronic excitations entering in the length scale for diffuse surface scattering.
While the QSE itself only produces a blueshift in energy of the LSPRs, the diffuse surface scattering mechanism gives to both energy and linewidth an oscillatory-damped behavior as a function of size, with characteristic lengths that depend on material parameters. Thus, the evolution of the LSPRs with size at the nanometer scale is very dependent on the relation of size to these lengths, which we illustrate with several examples. The variety of behaviors we find could be useful for designing plasmonic devices based on doped semiconductor nano structures having desired properties.

\end{abstract}

\maketitle

\newpage

\section{Introduction }

Strong interaction of metal nanosystems with light is the key element driving fundamental and applied research in plasmonics \cite{NMat_9_193_brongersma}. One of the key aspects that makes surface plasmons useful for e.g. cancer therapy \cite{AlexJMed_47-1-elsayed}, sensing \cite{Analyst_140_386_li}, solar harvesting \cite{JACS_135_5588_wang} and strong light-matter interactions \cite{PRL_114_157401_zengin} is their tuneability which enables resonances from the ultraviolet \cite{PCCP_15_5415_schatz} to infrared \cite{NatPhoton_6_409_stanley}. This is obtained by varying the size, geometry, materials, and arrangement of nanoparticles and their arrays \cite{plasmonics_maier, PRL_109_247401_tja}. Indeed, only through careful study of relationships between these parameters and the optical properties of plasmons did the previously mentioned and other applications become possible. 
Perhaps the most important parameter determining the resonances of localized surface plasmon resonances (LSPRs) is the nanoparticle size, a fact which can be inferred from the works of Mie \cite{AnnPhys_25_377_mie}. However, this classical treatment of the plasmon ceases to be applicable when the particle size reaches the single nanometer size range and the LSPR resonance deviates from the quasistatic prediction, either to the red or blue, depending on circumstances \cite{NJP_15_083044_carmina}. The need to elucidate these observations sparked extensive experimental and theoretical research which points towards quantum size and non-local surface effects as determining factors \cite{Nature_483_421_scholl, NatCom_5_3809_mortensen, PNAS_107_14530_peng, NL_9_3463_baida, NL_12_429_townsend, JPCL_1_2922_lerme, NJP_15_083044_carmina, OpEx_22_24994_carmina}.

Of the remaining parameters which dictate the behavior of LSPRs the plasma frequency $\omega_p$ of the metal plays also a key role. Derived from a fit of the Drude model to experimental data, $\hbar\omega_p$ ranges from a few eV (e.g. K with ca. 3.8~eV) to the teens of eV (e.g. Al with ca. 12-15 eV depending on the data) \cite{JPCC_113_3041_blaber, plasmonics_5_tja}, yielding LSPR resonances from the UV to the near-IR for spheres of the same size. The plasma frequency is a function of the carrier density which in metals is on the order of $10^{22}$--$10^{23}$ cm$^{-3}$, hence a reduction of the carrier density results in a smaller $\omega_p$. 
For example, doped semiconductors may have relatively high concentrations of electrons/holes in the conduction/valence bands, a prerequisite for supporting surface plasmons \cite{NMat_10_361_luther, NL_11_4706_buonsanti, PNAS_109_8834_naik}. Nanocrystals made from these materials through well controlled doping have carrier densities on the order of $n_c \simeq 0.5 \times 10^{21}- 5 \times 10^{21}$ cm$^{-3}$, a value two orders of magnitude lower than in typical metals \cite{NMat_10_361_luther}. Similar densities can also be achieved in synthesized colloidal metal oxide nanocrystals \cite{NL_11_4706_buonsanti}.
In common metals the classical density parameter $r_s=(\frac{3}{4 \pi n_c})^{\frac{1}{3}}$ expressed in units of the Bohr radius is 2--6, while in doped semiconductors it reaches 7--15. Even smaller electron densities have been recently achieved by photodoping of colloidal ZnO quantum dots (QDs), with $n_c \simeq 10^{20}$ cm$^{-3}$ or $r_s \simeq 25$ \cite{JPCL_4_3024_faucheux, JPCL_5_976_faucheaux, ACSNano_8_1065_schimpf}.
Materials with such dilute carrier densities exhibit an absorption band in the infrared rather than in the visible part of the electromagnetic spectrum and the energy of the LSPRs can be tuned by the doping level and the permittivity of the different host media in which the nanocrystal QDs are synthesized. 
LSPRs of doped semiconductors are very sensitive to changes in the electronic density. Therefore they can be used for detection of 
electronic proceses taking place at the nanoparticle which are important for redox chemistry.
Different ways of creating low carrier density QDs, as well as different aspects of their physical chemistry, are documented in the recent literature 
\cite{NMat_10_361_luther, NL_11_4706_buonsanti, PNAS_109_8834_naik, JPCL_4_3024_faucheux, JPCL_5_976_faucheaux, ACSNano_8_1065_schimpf,JPCL_5_1564_lounis}. 
It is clear that a better understanding and an accurate description of their optical properties is still needed to exploit potential applications of these new plasmonics systems.
From a fundamental, microscopic point of view, these systems can be considered as quantum objects since the wavelength of the carriers can be of the order of the crystal size, which makes them ideal systems for investigating quantum properties.
 
Optical properties of small particles are usually studied using the classical electromagnetic theory of Mie \cite{AnnPhys_25_377_mie} in which the electronic transport properties are described by a local (that is independent of spatial coordinates) dielectric function $\epsilon(\omega)$. In many cases $\epsilon(\omega)$ is just the bulk dielectric function of the material even though effects of quantization due to size (quantum size effect -- QSE) have recently been included \cite{Nature_483_421_scholl, ACSNano_8_1065_schimpf, JPCL_5_3112_prashant}.
For spheres of size in the nanometer scale, this is done by considering a system of free and independent electrons confined by an infinite potential barrier at the surface, then solving the Schr\"odinger equation and computing the dielectric function using the dipole approximation for the matrix elements. This approach thus concentrates on the discrete nature of the electronic states within the nanoparticle but does not take into account the effects of electron-electron interaction which are known to be important in the bulk and at the surface of the typical metals. 
In particular, the average potential a metal electron feels is not infinitely  abrupt at the surface but rather it is a smooth function of the spatial coordinate perpendicular to the surface, which allows the metallic wave functions to leak out of the metal \cite{ProgSurfSci_12_287_feibelman}. This is commonly known as the spill-out effect and, being a genuine surface property, strongly influences the LSPRs, both in energy and in width, of common metals not only for planar surfaces \cite{ProgSurfSci_12_287_feibelman, PRB_36_7378_liebsch, SS_247_302_tsuei} but also in nanosystems \cite{NL_12_429_townsend, PRL_110_263901_teperik, JPCC_118_16035_zhang, PRB_90_161407_zhang}.

A second surface mechanism, namely diffuse surface scattering, which affects LSPRs in any geometry originates from the existence of imperfections or roughness at a microscopic scale on the surface. Diffuse surface scattering causes, on average, omnidirectional reflection of electrons which arrive at a very rough surface from the bulk. This mechanism when applied to LSPRs contributes mainly to energy broadening since coherence between single scattering events is lost.
At high electron densities of typical metals the spill-out effect dominates over diffuse surface scattering \cite{ProgSurfSci_12_287_feibelman} but, as it was shown in Ref. \cite{JPCL_6_1847_carmina}, the spill-out effect decreases much more quickly with decreasing electron density than diffuse surface scattering. Therefore, at low enough electron densities, surface spill-out can be effectively neglected and diffuse scattering at the particle surface becomes the key mechanism contributing to blueshift and energy broadening of the plasmonic resonances in these systems.
One should notice that imperfections in the shape and morphology of ultra small particles, as well as the presence of absorbed molecules \cite{SurfSci_281_153_persson, JPCC_118_28075_mogensen}, are frequent and constitute the source of diffuse surface scattering.

In a previous publication \cite{JPCL_6_1847_carmina} we showed that, at ultra-low electron densities, diffuse scattering at the surface of nanospheres plays an important role in determining the resonance maximum and linewidth of LSPRs. We developed a detailed theoretical model that describes diffuse surface scattering and used it in conjunction with a dielectric function containing the QSE, giving results consistent with recent experimental results for photodoped ZnO \cite{ACSNano_8_1065_schimpf}. 
The purpose of the present work is to get deeper physical insight into the consequences of having diffuse scattering at the surface of low carrier density nanosystems. We identify and analyze the role that the different parameters appearing in the theory of the diffuse surface scattering effect play in giving the shape of the LSPRs. A key parameter is the length $R_0$, dependent on the density and effective mass of the charge carriers, which gives the strength of the QSE when compared to the particle radius $R$.
The other key parameter is the velocity of the electronic excitations. We obtain the length scale for diffuse surface scattering, $d_{\mathrm{dif}}$, with $d_\mathrm{dif}/R$ giving the strength of the effect. Other important lengths emerging from our analysis are the wavelength and the damping length of the oscillations appearing in the LSPRs which, in addition to the carrier density depend on other material parameters, and that can change completely the evolution of the plasmon width with size. 
We also quantify the relative contributions of QSE and diffuse surface scattering to the blueshift of the resonances. In addition to photodoped ZnO, we address other low carrier density nanosystems present in the literature. The variety of behaviors we find for the LSPRs could be useful for designing plasmonic systems having desired properties.

%%%%%%%%%%%%%%%%%%%%%%%%%%%%%%%%%%%%%%
\section{Theory}
%%%%%%%%%%%%%%%%%%%%%%%%%%%%%%%%%%%%%%

A microscopic formulation of the electromagnetic properties of bounded systems  requires a model for the behavior of charge carriers at the surface. This  will modify the electromagnetic fields near the surface with respect to the classical form. However, for comparison with most of the experiments, one only needs these fields integrated across the surface. In particular, for the case of  spheres of radius $R$, when the quasi-static limit $\frac{\omega}{c}R \ll 1$ is valid, $\omega$ being the frequency and $c$ the speed of light, the optical absorption cross section can be calculated as
\begin{equation}
\sigma(\omega,R)=4 \pi \frac{\omega}{c} \sqrt{\epsilon_m}\, \mathrm{Im}[ \alpha(\omega,R)],
\label{sigma-abs}
\end{equation}
with the polarizability of the sphere given by \cite{PS_26_113_apell,PRL_50_1316_apell}
\begin{equation}
\alpha(\omega,R)=\frac{R^3\left[\left(\epsilon(\omega)-\epsilon_m\right)\left(1-\frac{d_r(\omega,R)}{R}\right)+2\frac{d_{\theta}(\omega,R)}{R}\right]}
{\epsilon(\omega)+2\epsilon_m+2\left(\epsilon(\omega)-\epsilon_m\right) \frac{d_r(\omega,R)}{R}+2 \frac{d_{\theta}(\omega,R)}{R}}.
\label{alpha}
\end{equation}

In the above equations $\epsilon(\omega)$ is the classical, local permittivity of the metal, $\epsilon_m$ is the permittivity of the medium surrounding the sphere (assumed to be frequency independent) and the lengths $d_r(\omega,R)$ and $d_{\theta}(\omega,R)$ are defined as
\begin{subequations}
\begin{equation}
d_{r}(\omega,R)=\frac{1}{\frac{\epsilon(\omega)}{\epsilon_m}-1}\int_{0}^{\infty} dr \frac{E_{r}(r,\omega)-E_{r}^{cl}(r,\omega)}{E_{r}^{cl}(R,\omega)},
\label{dr}
\end{equation}
\begin{equation}
d_{\theta}(\omega,R)=\int_{0}^{\infty} dr \frac{r}{R} \frac{D_{\theta}(r,\omega)-D_{\theta}^{cl}(r,\omega)}{D_{\theta}^{cl}(R,\omega)},
\label{d-theta-def}
\end{equation}
\end{subequations}
respectively. In eqs. (\ref{dr}) and (\ref{d-theta-def}) $r$ is the radial coordinate, $E_{r}$ and $D_{\theta}$ denote the normal to the surface component of the electric field vector $\mathbf{E}$ and the parallel to the surface component of the displacement vector $\mathbf{D}$, respectively.  
$E_{r}^{cl}$ and $D_{\theta}^{cl}$ are their classical counterparts for a model where there is an abrupt change from $\epsilon(\omega)$ to $\epsilon_m$ at $r=R$.
Note that $E_{r}$ and $D_{\theta}$ differ from  $E_{r}^{cl}$ and $D_{\theta}^{cl}$ only in the region near the surface, recovering the classical results of the Mie theory if $d_r=0$ and $d_{\theta}=0$ in eq.~(\ref{alpha}).  The lengths $d_r$ and $d_{\theta}$ when compared to the size $R$ give, respectively, the strength of the spill-out and diffuse surface scattering effects. This can be understood from the following argument. $d_{r}$ can be related to the electronic charge density, $\delta\rho$, induced at the surface by any external perturbation as
\begin{equation}
\frac{d_r(\omega)}{R}=\frac{ \int_{0}^{\infty} dr\; r(R-r)\delta\rho(r,\omega)}
{ \int_{0}^{\infty} dr\; r^2 \delta\rho(r,\omega)}.
\label{dr-rho}
\end{equation}
Notice in eq.~(\ref{dr-rho}) that  $d_{r}=0$ in the classical case, where the induced charge density is a $\delta$-function at $r=R$, hence the length $d_{r}$ describes the spill-out effect. Now consider  an ideal perfectly flat surface. Electron scattering at such a surface does not change the parallel component of the electron momentum and the parallel current is conserved. Since $E_{\theta}$ is a continuous function across the surface, $D_{\theta}$ is not modified with respect to its classical form and $d_{\theta}=0$.  However, at a very rough surface (on a microscopic scale) an electron can, on the average, be scattered back in any direction and the parallel current is not conserved. Consequently, $d_{\theta} \neq 0$. 
 
We should stress that $d_r(\omega,R)$ and $d_{\theta}(\omega,R)$ are complex surface response functions whose real and imaginary parts are not independent but related by Kramers-Kronig relations \cite{PRB_27_6058_persson}. From the structure of eq.~(\ref{alpha}) it can be  readily seen that their real parts contribute mainly to the energy shift while the imaginary parts contribute to the change in width of the plasmon resonances with particle size. Then position and width of the LSPRs are, in general, related magnitudes. A common simple model for describing surface scattering consists of defining an effective damping rate as \cite{ZPhys_224_307_kreibig}
\begin{equation}
\gamma_{eff}\equiv\frac{1}{\tau_{eff}}=\frac{1}{\tau_b}+A\frac{v_F}{R},
\label{tau-eff}  
\end{equation}
where $\tau_b$ is the bulk relaxation time due to scattering with phonons and impurities, $v_F$ is the Fermi velocity and $A$ is a constant of the order of unity which is usually taken as an adjustable parameter. The effective damping rate is then introduced in a Drude dielectric function of the form  $\epsilon(\omega)=\epsilon_{\infty}-\frac{\omega_p^2}{\omega^2+i\omega \gamma_{eff}}$,
where $\omega_p=\sqrt { \frac{n_c e^2}{\epsilon_0 m^{*}m_e}}$ is the plasma frequency,  $\epsilon_{\infty}$ is the high-frequency permittivity of the material, $\epsilon_0$ is the permittivity of the free space, and $m^{*}$ is the effective mass of the charge carrier in units of the electron mass $m_e$.
The calculation of the absorption cross section using this Drude dielectric function in the classical Mie theory produces LSPRs whose  position is fixed at the Drude value  $\omega_D=\omega_p/\sqrt{\epsilon_{\infty}+2\epsilon_m}$ independent of $R$ and only the linewidth scales linearly with $1/R$. This is because in this classical model the size of the system only limits the mean free path of carriers. Therefore this model can only approximate true surface scattering in cases in which the mechanism slightly affects the plasmon energy, which only occurs in especial circumstances, as we will see below. 

In this work we use a theory for diffuse surface scattering that was first designed for planar surfaces \cite{JPhysFr_38_863_flores, PS_22_155_monreal, JPhysFr_43_901_monreal} and then extended to spheres \cite{PRB_32_7878_deAndres}.  As a detailed account of it is given in Ref. \cite{JPCL_6_1847_carmina} and in the accompanying Supplementary Information, we only reproduce here the basic ingredients.
The theory embeds a real sphere in an infinite, fictitious medium, both characterized by exactly the same dielectric functions. Then, a constitutive relation for the polarization $\mathbf{P}_f(\mathbf{r}, \omega)$ caused by free charges inside the real sphere, is written as 
\begin{equation}
\frac{1}{\epsilon_0} \mathbf{P}_f(\mathbf{r}, \omega)=\int d^3 \mathbf{r'} [\bm{\epsilon}(\mathbf{r}-\mathbf{r'}, \omega)- 
\epsilon_{\infty} \mathbf{I} \delta(\mathbf{r}-\mathbf{r'})] \cdot \mathbf{E}^{M}(\mathbf{r'}, \omega),
\label{P-general}
\end{equation}
where  $\mathbf{E}^{M}$ is the electric field vector in the infinite medium, which for $|\mathbf{r}|< R$ is the actual electric field inside the sphere. $\bm{\epsilon}$ is the dielectric tensor of the medium and the integral extends to the whole space. In this integral, the region of space  $|\mathbf{r'}|< R$ describes excitations produced at a point $\mathbf{r'}$ inside the sphere that propagate directly to the point $\mathbf{r}$.
The fictitious region occupying $|\mathbf{r'}|> R$ simulates excitations which arrive at $\mathbf{r}$ after having been reflected at the surface. Therefore, the surface properties are mimicked by values of the electric field $\mathbf{E}^{M}$ in the fictitious region of the infinite medium. We want the sphere surface to reflect electrons completely at random meaning that, on the average, no excitation originating at the surface will arrive at $\mathbf{r}$. These conditions require that an equivalent electric field $\mathbf{E}^{M}$ in eq. (\ref{P-general}) be zero outside the sphere. Hence, the problem consists of constructing an electric field of the form
\begin{equation}
\mathbf{E}^{M}(\mathbf{r}, \omega)=\left\{
\begin{array}{ll}
\mathbf{E}_{sphere}(\mathbf{r}, \omega) & \mathrm{for}\;\; |\mathbf{r}|< R\nonumber \\
0 & \mathrm{for}\;\; |\mathbf{r}|> R
\end{array} \right.
\end{equation}
which satisfies the Maxwell equations. To do so, one needs a model for the dielectric tensor, $\bm{\epsilon}$, of the medium.

An essential requirement for a theory of diffuse surface scattering is the use of a non-local transverse dielectric function, in the same way that a non-local longitudinal dielectric function is required to describe the spill-out effect. The simplest possible form for a non-local transverse dielectric function is an analogue to the familiar hydrodynamical model for the longitudinal dielectric function.
In our approximation $\bm{\epsilon}$ only depends on spatial coordinates through the difference $\mathbf{r}-\mathbf{r'}$, it is thus convenient to Fourier-transform the permittivity to momentum space where our non-local transverse dielectric function reads
 \begin{equation}
\epsilon_T(\mathbf{k},\omega)=\epsilon_{\infty}-\frac{\omega_p^2}{\omega^2-\Delta^2+i\omega\gamma_b-\beta_T ^2 k^2},
\label{epsilonT}
\end{equation}
where $\mathbf{k}$ is a wave vector and $\beta_T$ is a constant proportional to the Fermi velocity $v_F$. Here, the hydrodynamical-like dielectric function has been implemented to include QSE by means of the energy gap  $\Delta=\omega_p\frac{R_0}{R}$ (see Ref. \cite{NJP_15_083044_carmina}) with $R_{0}=\sqrt{\frac{3\pi a_0}{4m^{*} k_F}}$ \cite{SovPhysJETP_21_940_gorkov},  $k_F=(3 \pi^2 n_e)^\frac{1}{3}$ being the Fermi wave vector. The energy gap $\Delta$ (or the length $R_0$) is the first key parameter of our theory.

Since we can neglect spill-out effects \cite{JPCL_6_1847_carmina} (low electron density materials), the longitudinal dielectric function of the sphere is a local one. This choice yields $d_r=0$. The longitudinal and the transverse dielectric functions have to be equal in the $\mathbf{k}=0$ limit, therefore a good approximation for $\epsilon_L$ is
\begin{equation}
\epsilon_L(\omega)=\epsilon_{\infty}-\frac{\omega_p^2}{\omega^2-\Delta^2+i\omega\gamma_b}.
\label{epsilonL}
\end{equation}

Having specified the dielectric response of the medium the electromagnetic normal modes of the electronic system are known. The longitudinal modes, which are solutions of the equation $\epsilon_L(\mathbf{k},\omega)=0$, are absent because $\epsilon_L$ is a local dielectric function. The transverse modes are solutions of the equation
\begin{equation}
k^2-\frac{\omega^2}{c^2} \epsilon_T(k, \omega)=0.
\end{equation}
For the model of eq.~(\ref{epsilonT}) we have two transverse modes given by
\begin{equation}
T_{1,2}^2=\frac{1}{2}\left[\left(t^2+\frac{\omega^2}{c^2}\epsilon_{\infty}\right) \pm \sqrt{\left(t^2+\frac{\omega^2}{c^2}\epsilon_{\infty}\right)^2
-4\frac{\omega^2}{c^2}\epsilon_{\infty}S^2}\right] ,
\label{T12}
\end{equation}
where $t$ and $S$ are the pole and the zero of $\epsilon_T$, respectively, and are given by
\begin{equation}
t^2=\frac{\omega^2 -\Delta^2 +i\omega\gamma_b}{\beta_T^2},
\label{eq-t}
\end{equation}
and
\begin{equation}
S^2=t^2-\frac {\omega_p^2}{\epsilon_{\infty} \beta_T^2}.
\end{equation} 

Since $\beta_T$ is of the order of $v_F$, and therefore $\frac{\beta_T}{c} \ll 1$, it can be readily seen from eq.~(\ref{T12}) that, to the order  $\beta_T/c$, $T_1 \simeq t$ and $T_2 \simeq k_t=\frac{\omega}{c}\sqrt{\epsilon_T(k=0,\omega)}$. Hence the present approximation contains the usual polariton mode propagating with wave vector $k_t$ and one transverse electron-hole pair propagating with wave vector $t$ given by eq.~(\ref{eq-t}). 
Then our simple form of $\epsilon_T$ substitutes the whole spectrum of electronic excitations by just a single electron-hole pair whose energy is proportional to momentum as $\omega \simeq \beta_T t $, $\beta_T$ being the velocity of the pair. This means that an appropriate value for $\beta_T$ could be an average value over the whole spectrum. 
For large metallic systems, $\beta_T=\frac{v_F}{\sqrt 5}$ in order to fit the low frequency limit of the Lindhard dielectric function even though other choices can be found in the literature. The velocity  $\beta_T$ is the other key parameter of the theory ($R_0$ being the first). 

Once the electromagnetic field is found, the absorption cross-section of the sphere can be obtained and  compared  with the form of eqs. (\ref{sigma-abs}) and (\ref{alpha}) in the quasi-static limit. This allows us to obtain the  length $d_{\theta}(\omega,R)$ as
\begin{equation}
\frac{d_{\theta}(\omega, R)}{R}=-i(\epsilon-\epsilon_{\infty})(tR)h_{1}^{(1)}(tR)j_{1}(tR),
\label{d-theta}
\end{equation}
where $j_1$ and $h_{1}^{(1)}$ are the spherical Bessel functions and $\epsilon(\omega)=\epsilon_T(k=0,\omega)$. We have checked numerically that eq.~(\ref{d-theta}) fulfills Kramers-Kronig relations, as it should. Eq.~(\ref{d-theta}) is the most important result of our theory for diffuse surface scattering.

Finally, an effective size-dependent dielectric function for a sphere, $\tilde\epsilon(\omega, R)$, is  derived as
\begin{equation}
\tilde\epsilon(\omega, R)\equiv\epsilon(\omega)+2 \frac{d_{\theta}(\omega, R)}{R}   = \epsilon(\omega)+\epsilon_{\theta}(\omega,R),
\end{equation}
where we have defined
\begin{equation}
\epsilon_{\theta}(\omega,R)\equiv -2 i(\epsilon(\omega)-\epsilon_{\infty})(tR)h_{1}^{(1)}(tR)j_{1}(tR).
\label{epsilon-theta}
\end{equation}
We call $\epsilon_{\theta}(\omega,R)$ the surface correction to the dielectric function due to diffuse scattering. It is also the basic function for analyzing results presented in the next section. To this end, we use large-argument asymptotic expansions of the Bessel functions, which are good approximations for the values of $R$ used here, to yield
\begin{equation}
\epsilon_{\theta}(\omega, R) \approx i \frac{\omega_p^2}{(\omega ^2-\Delta^2+i\omega\gamma_b)^{\frac{3}{2}}}
\frac{\beta_T}{R}\left[1+e^{i 2 t R}\right],
\label{epsilon-theta-asym}
\end{equation}
which exhibits the proportionality between $\epsilon_{\theta}$ and the parameter $\beta_T$. Defining the complex variable $z$ as
\begin{equation}
z\equiv tR=\frac{\omega_D R}{\beta_T}\sqrt{\tilde\omega^2-\tilde\Delta^2+i \tilde\omega \tilde\gamma_b},
\label{z}
\end{equation}
where we have used the reduced magnitudes $\tilde\omega\equiv\frac{\omega}{\omega_D}$, $\tilde\Delta\equiv\frac{\Delta}{\omega_D}=\sqrt{\epsilon_{\infty}+2\epsilon_m} \frac{R_0}{R}$ and $\tilde\gamma_b\equiv\frac{\gamma_b}{\omega_D}$ 
($\omega_D=\omega_{p}/\sqrt{\epsilon_{\infty}+2\epsilon_m}$ being the Drude frecuency),  eq.~(\ref{epsilon-theta-asym}) reads
\begin{equation}
\epsilon_{\theta}(\omega, R) = i \frac{\beta_T}{\omega_D R} \frac{\epsilon_{\infty}+2\epsilon_m}{(\tilde\omega ^2-\tilde\Delta^2+i\tilde\omega\tilde\gamma_b)^{\frac{3}{2}}}
\left[1+e^{i 2 z}\right]. 
\label{epsilon-theta-basic}
\end{equation}
Furthermore, since $\gamma_b \leq \omega_D$, $z$ in eq.~(\ref{z}) can be approximated as
\begin{equation}
z \approx \frac{\omega_D R}{\beta_T}\left[ \sqrt{\tilde\omega^2-\tilde\Delta^2}+\frac{i}{2} \tilde\gamma_b \frac{\tilde\omega}{\sqrt{\tilde\omega^2-\tilde\Delta^2}}\right].
\label{z-approx}
\end{equation}

When eq.~(\ref{z-approx}) is substituted into eq.~(\ref{epsilon-theta-basic}) we find different behaviors of $\epsilon_{\theta}$ depending on $R$. The region of large values of $R$ where QSEs do not impact surface scattering  is defined as $\tilde\Delta \ll 1$. In this region  $\tilde\omega \approx 1$, and eq.~(\ref{z-approx}) reads
\begin{equation}
z \approx \frac{\omega_D R}{\beta_T} + \frac{i}{2} \frac{\gamma_b R}{\beta_T}.
\label{z-approx-approx}
\end{equation}
Then, $\epsilon_{\theta}$ in eq.~(\ref{epsilon-theta-basic}) is practically independent of $\tilde\omega$ and can be approximated as

\begin{equation}
\epsilon_{\theta}\approx i \frac{\beta_T}{\omega_D R}(\epsilon_{\infty}+2\epsilon_m)\left[1+e^{i 2 z}\right], 
\label{epsilon-theta-long}
\end{equation}
with $z$ given by eq.~(\ref{z-approx-approx}). Equation~(\ref{epsilon-theta-long}) allows us to define the length scale for diffuse surface scattering as
\begin{equation}
d_\mathrm{dif}\equiv \frac{\beta_T}{\omega_D }(\epsilon_{\infty}+2\epsilon_m)=\frac{\beta_T}{\omega_p }(\epsilon_{\infty}+2\epsilon_m)^{\frac{3}{2}}.
\label{d-diffuse}
\end{equation}
Moreover, eq~(\ref{epsilon-theta-long}) presents damped oscillations as a function of $R$, a characteristic that will be translated to the energy and width of the LSPRs. The oscillations are of wave length $\lambda_o \simeq \pi \frac{\beta_T}{\omega_D}$ and are damped within a typical length $\lambda_d=\frac{\beta_T}{\gamma_b}$.  For $R$ large enough so that $\frac{R}{\lambda_d} \gg 1$, the oscillations are fully damped and the sphere surface behaves as a planar surface for scattering of electrons \cite{SSComm_52_971_apell}. In this case 
\begin{equation}
\epsilon_{\theta}\approx i \frac{d_\mathrm{dif}}{R}
\end{equation}
and $\epsilon_{\theta}$ has only an imaginary part. Then, the effect of diffuse surface scattering is to broaden the LSPRs for decreasing $R$ linearly with $1/R$ with a slope proportional to $\beta_T$ (or $v_F$) without moving their position in energy, which are just the results of the simple model of eq.~(\ref{tau-eff}). In the intermediate region, $\frac{R}{\lambda_d} \approx  1$, the oscillations of $\epsilon_{\theta}$ show up in the LSPRs and the wave length $\lambda_o$ can be used for an experimental determination of the parameter $\beta_T$.
Finally another region is that of small $R$, $\tilde\Delta \simeq 1$, where QSE have an impact on diffuse surface scattering and $\epsilon_{\theta}(\omega,R)$ depends on both of its variables. At shorter $R$, eq.~(\ref{epsilon-theta-basic}) is not such a good approximation to $\epsilon_{\theta}$ and a numerical evaluation of eq.~(\ref{epsilon-theta}) is necessary. Examples of the different regimes are presented in the next section.

We have just stressed that QSEs affect the surface scattering properties at small sizes. However, QSEs appear in our theory even if we neglect diffuse surface scattering by making $\beta_T=0$. In this case we are left with a local dielectric function of the Lorentz type: $\epsilon(\omega)=\epsilon_{\infty}-\omega_p^2(\omega^2- \Delta^2+i\omega \gamma_{b})^{-1}$, which blueshifts the resonance energy relative to its Drude value as
\begin{equation}
\frac{\omega_{QSE}}{\omega_D}=\sqrt{1+\tilde\Delta^2},
\label{omega-QSE}
\end{equation}
while the linewidth equals $\gamma_b$ independent of $R$. Therefore, in our general theory with both QSEs and diffuse surface scattering present, the width of the LSPRs is larger than $\gamma_b$ as a consequence of the later, even though the amount of increase may be very dependent on $\Delta$ for small $R$. However, as both effects contribute independently to the blueshift of the LSPR energy, an estimation of their relative contributions is given in the next section. At this point let us comment that if the surface plasmon energy is dominated by QSEs, then
$\tilde \omega \simeq \frac{\omega_{QSE}}{\omega_D}$, $\tilde \omega^2-\tilde \Delta^2 \simeq 1$, and the approximate value of $z$ to be 
substituted in eq~(\ref{epsilon-theta-long}) is
\begin{equation}
z \approx \frac{\omega_D R}{\beta_T} + \frac{i}{2} \frac{\gamma_b R \sqrt{1+\tilde\Delta^2}}{\beta_T}.
\label{z-approx-delta}
\end{equation}
Therefore, increasing $R_0$ makes the damping length of the oscillations to effectively decrease. Hence strong QSEs tend to decrease the plasmon width.
We also notice in eqs~(\ref{epsilon-theta-basic}), (\ref{z-approx}) and (\ref{omega-QSE}) that it is $\tilde R_0=R_0 \sqrt{\epsilon_{\infty}+2\epsilon_m}$ rather than $R_0$ itself that is the length controlling the strength of the QSE.

\section{Results and discussion}

We calculate the absorption cross section (eq \ref{sigma-abs})\cite{JPCL_6_1847_carmina} to obtain the position, $\omega_R$, and the width, $\Gamma_R$, of the LSPRs. The width is defined as the full width at half maximum of the absorption curve. We first show calculations for the case of ultralow-electron-density photodoped ZnO nanocrystals in toluene, using the experimental values of the parameters: $\epsilon_{\infty}=3.72$, $\epsilon_m=2.25$, $m^{*}=0.28$ and $\gamma_b=$0.1 eV \cite{ACSNano_8_1065_schimpf}.
With these values, we use the lower limit of the experimental electron density, $n_e=1 \times 10^{20}$ cm$^{-3}$, to get the experimental surface plasmon energies at the largest radii. Then, $r_s=25.3$, the Fermi velocity $v_F=0.59 \times 10^{6}$ m s$^{-1}$ and $\omega_p=\sqrt { \frac{n_e e^2}{\epsilon_0 m^{*}m_e}}= 0.70$~eV.
To illustrate the different behaviors of $\epsilon_{\theta}(\omega, R)$, $\beta_T$ and $R_0$ are varied from their reference values, $\beta_{T}^{(r)}=v_F/\sqrt 5$ and $R_{0}^{(r)}=\sqrt{\frac{3\pi a_0}{4m^{*} k_F}}= 0.56$~nm, by less than 40$\%$. 
The crystal size is in the nanometer range with 1.5 nm $< R < $ 12 nm. We note that the scale length for diffuse surface scattering $d_\mathrm{dif}^{(r)}=5.85$~nm is in the middle of the range so we can expect substantial effects in this system. Moreover, since $\tilde R_{0}^{(r)}= 1.6$~nm, QSEs have an impact on $\Gamma_R$ and $\omega_R$ at short radii.
 
\begin{figure}
\centering
\includegraphics{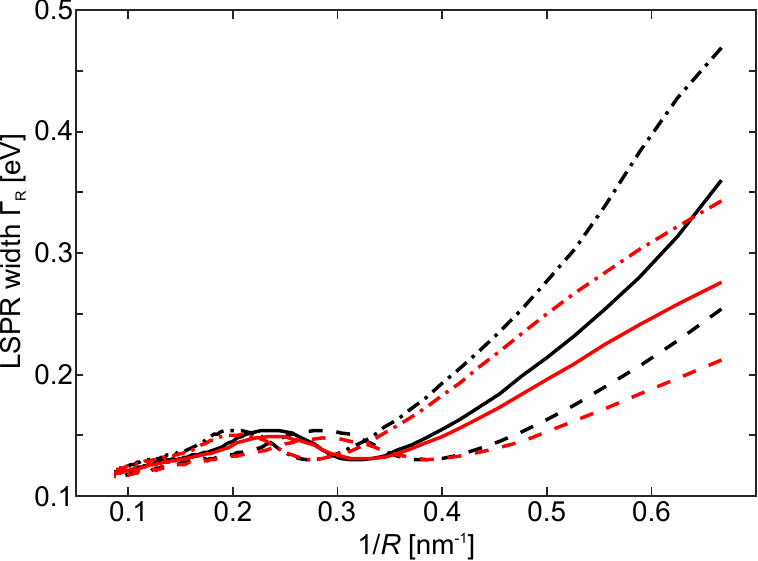}
\caption{Width of the LSPRs of ultralow-electron-density ZnO spheres in toluene as a function of $R^{-1}$ for $R_0=0$ (black lines) and $R_0=1.2 R_{0}^{(r)}$ (red lines) and different values of $\beta_T$, $\beta_T=\beta_{T}^{(r)}$ (dashed lines), $\beta_T=1.2 \beta_{T}^{(r)}$ (continuous lines) and $\beta_T= 1.4 \beta_{T}^{(r)}$ (dot-dashed lines). Even though $\Gamma_R$ is very sensitive to the values of both parameters at short radii,
it shows a clear decrease with increasing strength of the QSE.}
\label{fig-width}
\end{figure}

Figure~\ref{fig-width} shows the LSPR width $\Gamma_R$ as a function of $R^{-1}$ for $R_0=0$ (QSEs neglected) and $R_0=1.2 R_{0}^{(r)}$ and three values of $\beta_T$. The different behaviors of $\epsilon_{\theta}$ described above show up clearly in the width. We see the region of very large values of $R$ where the width scales linearly with $R^{-1}$ with a slope proportional to $\beta_T$. 
Then, as discussed above, the oscillatory region appears ($0.2 \lesssim R^{-1} \lesssim 0.4$ nm$^{-1}$) where QSEs are not important and the width depends weakly on $R_0$.  In this region $\epsilon_{\theta}$ depends on $R$ only trough the ratio $R/\beta_T$ so that a change in $\beta_T$ by a factor $f$ is completely equivalent to rescale R by the same factor (see eqs. (\ref{z-approx-approx}-\ref{epsilon-theta-long})), a characteristic translated to the width.
Finally, for $R^{-1} \gtrsim 0.5$ nm$^{-1}$, QSEs come into play. Notice that in this region $\Gamma_R$ is very sensitive to the values of both parameters $\beta_T$ and $R_0$, even though it follows the trend of increasing width with increasing $\beta_T$ for fixed $R_0$. For fixed $\beta_T$, the width decreases with increasing $R_0$ following the increase of $\mathrm{Im}[z]$ in eq.~(\ref{z-approx-delta}), as commented above. Hence strong QSEs tend to decrease the plasmon width.
  
\begin{figure}
\centering
\includegraphics{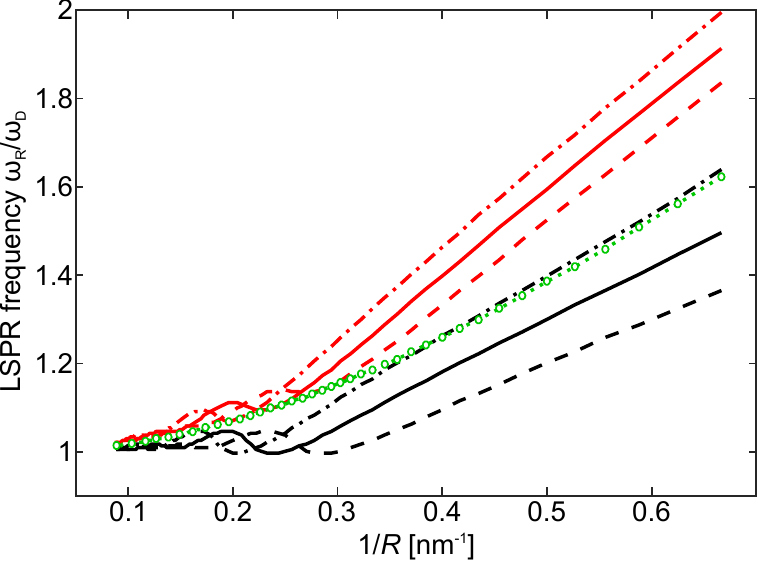}
\caption{The energy of the LSPRs (relative to the Drude value) of ultralow-electron-density ZnO spheres in toluene as a function of $R^{-1}$ for $R_0=0$ (black lines) and $R_0=1.2 R_{0}^{(r)}$ (red lines) and different values of $\beta_T$, $\beta_T=\beta_{T}^{(r)}$ (dashed lines), $\beta_T=1.2 \beta_{T}^{(r)}$ (continuous lines) and $\beta_T= 1.4 \beta_{T}^{(r)}$ (dot-dashed lines). The values of $\omega_{QSE}/\omega_D$ for $R_0=1.2 R_{0}^{(r)}$ are shown by green dots.}
\label{fig-wR}
\end{figure}

Figure~\ref{fig-wR} displays $\omega_R$ normalized to its Drude value for the same values of the parameters as in Figure~\ref{fig-width}. The resonance position is strongly blueshifted even in the absence of QSEs ($R_0=0$) and behaves approximately as the width.  However, for $R^{-1}\gtrsim 0.3$ nm$^{-1}$, it shows a remarkable linear scaling with $R^{-1}$ with a slope increasing with both $R_0$ and $\beta_T$. As reference for the importance of the QSEs alone, the results of eq.~(\ref{omega-QSE}) for $R_0=1.2 R_{0}^{(r)}$ are shown by dots. At large radii ($R^{-1}\lesssim 0.25-0.35$~nm$^{-1}$) the surface plasmon energy is dominated by QSEs. The radius at which diffuse surface scattering becomes important is determined by $\beta_{T}$ -- the larger $\beta_{T}$ is, the larger is the critical radius below which diffuse scattering is visible.

\begin{figure}
\centering
\includegraphics{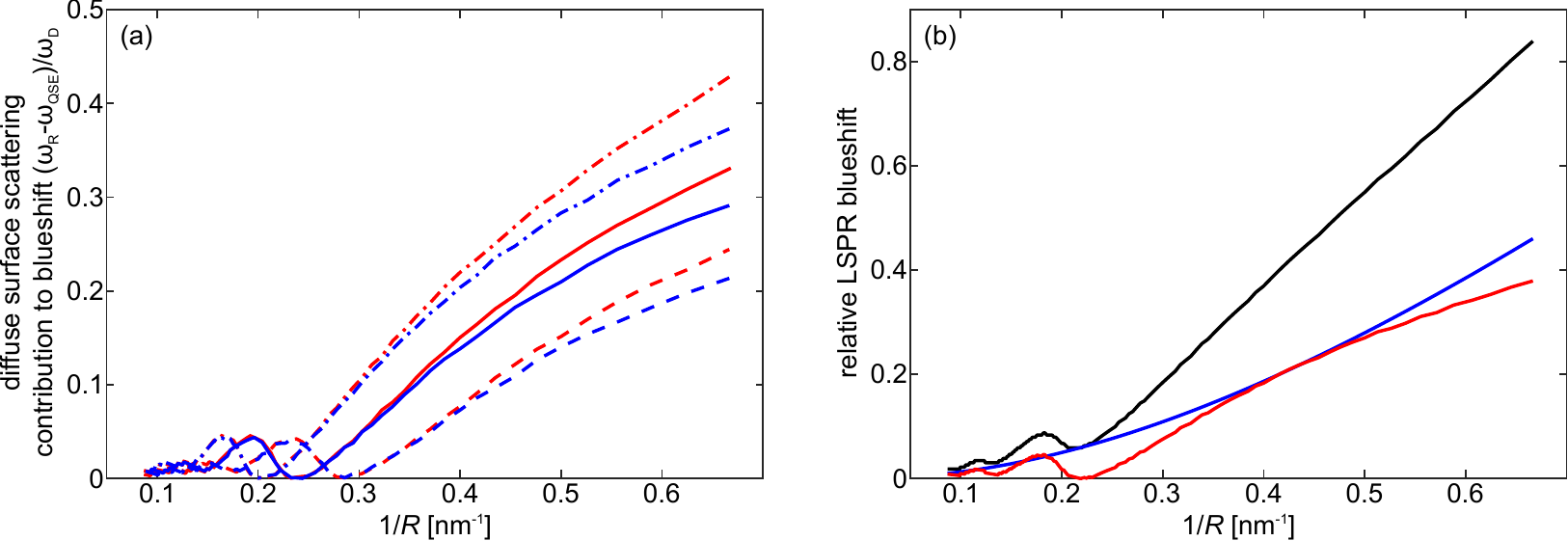}
\caption{(a) Contribution of diffuse surface scattering to the blueshift 
and of ultralow-electron-density ZnO spheres in toluene as a function of $R^{-1}$ for $R_0=R_{0}^{(r)}$ (red lines) and $R_0=1.2 R_{0}^{(r)}$ (blue lines) and different values of $\beta_T$, $\beta_T=\beta_{T}^{(r)}$ (dashed lines), $\beta_T=1.2 \beta_{T}^{(r)}$ (continuous lines) and $\beta_T= 1.4 \beta_{T}^{(r)}$ (dot-dashed lines).
(b) The relative blueshift in energy of the LSPRs of ultralow electron density ZnO spheres as a function of $R^{-1}$ (black line) is obtained as the sum of the contribution of QSEs (blue line) and the contribution of diffuse surface scattering (red line), for $R_0=R_0^{(r)}$ and $\beta_T=1.3 \beta_{T}^{(r)}$ the values of these parameters that give a good account of the experimental values of the energy and width of the LSPRs in Ref. \cite{ACSNano_8_1065_schimpf}.
The relative blueshift reaches 80$\%$ at the shortest radii with both, QSEs and diffuse surface scattering contributing nearly the same amount.}
\label{fig-wdiff}
\end{figure}

We want to quantify the relative contributions of QSEs and diffuse surface scattering to the energy of the LSPRs. This is not always strictly possible since, as we said above, QSEs modify the diffuse surface response $\epsilon_{\theta}$ in a non-trivial way. However, guided by the results of Figure \ref{fig-wR}, we estimate the effect of quantization due to size using the function $\omega_{QSE}/\omega_D$ of eq.~(\ref{omega-QSE}), then 
$\omega_{\mathrm{dif}}/\omega_D \equiv (\omega_R-\omega_{QSE})/\omega_D$ defines our estimation of the effect of diffuse surface scattering. This function is plotted in Figure \ref{fig-wdiff}a for three values of $\beta_T$  and two values of $R_0$, showing that it depends strongly on $\beta_T$ being not so dependent on $R_0$, so it gives a good estimate of the effect of diffuse surface scattering. 

We now define the total blueshift of the resonance energy, relative to $\omega_D$, as $(\omega_R-\omega_D)/\omega_D$, the relative contribution to the energy blueshift of QSEs as $(\omega_{QSE}-\omega_D)/\omega_D$, and the contribution of diffuse surface scatteing is then $\omega_{\mathrm{dif}}/\omega_D=(\omega_R-\omega_{QSE})/\omega_D$.
These contributions are shown in Figure~\ref{fig-wdiff}b for the parameters that give a good account of the experimental results of Ref. \cite{ACSNano_8_1065_schimpf}, $R_0=R_0^{(r)}$ and $\beta_T=1.3 \beta_{T}^{(r)}$ \cite{JPCL_6_1847_carmina}. Notice that the blueshift of the resonance can be as large as 80$\%$ of the Drude value. One sees that both QSEs and diffuse surface scattering contribute nearly the same to the experimental blueshift at all radii such that $R^{-1}\gtrsim 0.3$ nm$^{-1}$.
 
\begin{figure}
\centering
\includegraphics{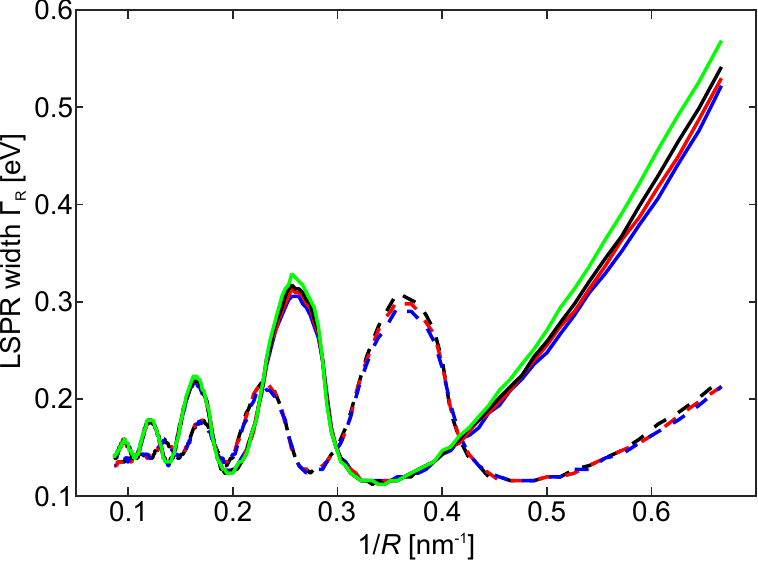}
\caption{Width of the LSPRs of low-electron-density ZnO as a function of $R^{-1}$ for two values of $\beta_T$, $\beta_T=\beta_{T}^{(r)}$ (dashed lines), $\beta_T=1.4 \beta_{T}^{(r)}$ (continuous lines) and $R_0=0$ (green lines), $R_0=R_{0}^{(r)}$ (black lines), $R_0=1.2 R_{0}^{(r)}$ (red lines), and $R_0=1.4 R_{0}^{(r)}$ (blue lines). In this system the plasmon width has an oscillatory-damped behavior in the whole nanometer range of radii which is independent of
the QSE.}
\label{fig-width-n10}
\end{figure}

In the experiments of Schimpf et al.\cite{ACSNano_8_1065_schimpf} the electron density was estimated to be $n_e=(1.4 \pm 0.4) \times 10^{20}$ cm$^{-3}$ and we used the lower limit to reproduce the experimental values of $\omega_R$ at large $R$. However, we have checked that the behavior of the LSPRs just described does not change qualitatively when changing the electronic density within the experimental uncertainty.
A different pattern would have been encountered if the electronic density could have been increased by a factor of 10, to $n_e=1 \times 10^{21}$ cm$^{-3}$, without changing the rest of the material parameters. We call this system low-electron-density ZnO in toluene for which $r_s=11.7$,  $v_F=1.23 \times 10^{6}$~m s$^{-1}$  $\omega_p=2.22$~eV, $R_0^{(r)}=0.38$~nm, $\tilde R_0^{(r)}=1.09$~nm, and  $d_\mathrm{dif}^{(r)}=3.96$~nm.

\begin{figure}
\centering
\includegraphics{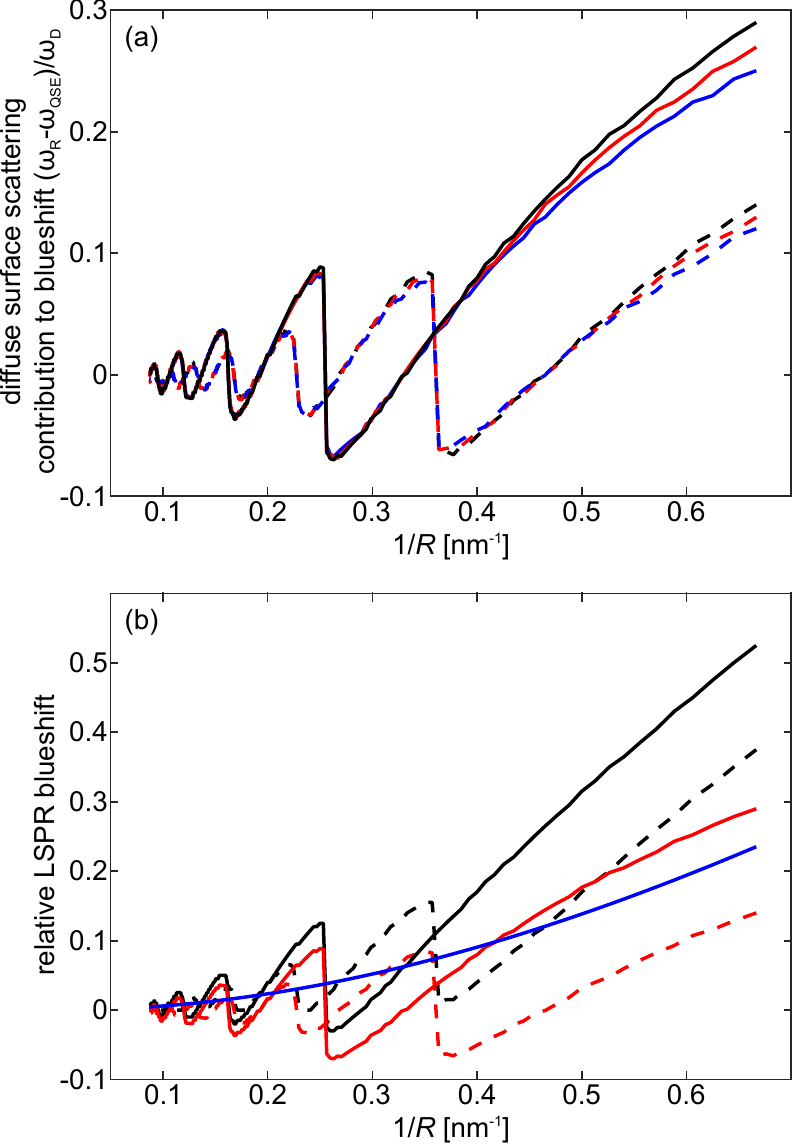}
\caption{(a) Contribution of diffuse surface scattering to the blueshift of low-electron-density ZnO spheres as a function of $R^{-1}$ for the same values of the parameters as in Figure \ref{fig-width-n10}. 
(b) The relative blueshift in energy of the LSPRs of low-electron-density ZnO spheres as a function of $R^{-1}$ (solid lines) is obtained as the sum of the contribution of QSE (blue line) and the contribution of diffuse surface scattering (dashed lines).
 $R_0=R_{0}^{(r)}$ and two values of $\beta_T$, $\beta_T=\beta_{T}^{(r)}$ (red lines), $\beta_T=1.4 \beta_{T}^{(r)}$ (black lines).
Even though the relative blueshift is smaller than for ultralow-electron-density ZnO, both effects contribute nearly the same also in this case.
The saw-tooth oscillations of the diffuse scattering contribution can change the sign of the energy shift from blue to red.}
\label{fig-wdiff-n10}
\end{figure} 

Figure~\ref{fig-width-n10} shows $\Gamma_R$ of low-electron-density ZnO as function of $R^{-1}$ for several values of the parameters $R_0$ and $\beta_T$. Two differences with the previous case are noticeable. First, the decrease in $\tilde R_0^{(r)}$ causes that QSEs do not impact diffuse surface scattering. Therefore they are actually independent effects and the respective contributions can be added. 
Second, the oscillatory-damped behavior of $\epsilon_{\theta}$ occurs in the whole nanometer range of $R$ and the oscillations are less damped as a consequence of having a smaller wave length and a larger damping length than in the experimental ultralow-electron-density system. The wave length of the oscillations is $\lambda_o \simeq 1.5-2$~nm and the effect could be detected in samples with a dispersion in sizes smaller than 1 nm, which seems experimentally feasible. 
The rescaling of $R$ with $\beta_T$ in the whole range of radii is also apparent in the Figure.
Figure~\ref{fig-wdiff-n10}a shows $\omega_{\mathrm{dif}}/\omega_D$ for the same values of the parameters as in Figure~\ref{fig-width-n10}. It reveals 
a saw-tooth-damped behavior and also the rescaling of $R$ with $\beta_T$. Notice that 
the contribution of diffuse surface scattering to the energy shift at short radii is smaller than the one shown in Figure~\ref{fig-wdiff}a because    
the value of $d_\mathrm{dif}^{(r)}$ is smaller than for ultralow-electron-density ZnO. 
The relative contributions of QSE and diffuse surface scattering to the energy shift of the LSPRs are displayed in Figure~\ref{fig-wdiff-n10}b. Even tough the total relative energy shift is not so large as for ultralow-electron-density ZnO (Figure~\ref{fig-wdiff}b), because of the smaller values of both
$d_\mathrm{dif}^{(r)}$ and 
$\tilde R_0^{(r)}$, the contributions of both effects can be similar also in this case. It is interesting to note that the oscillations caused by diffuse surface scattering can change the sign of the energy shift from blue to red. 

\begin{figure}
\centering
\includegraphics{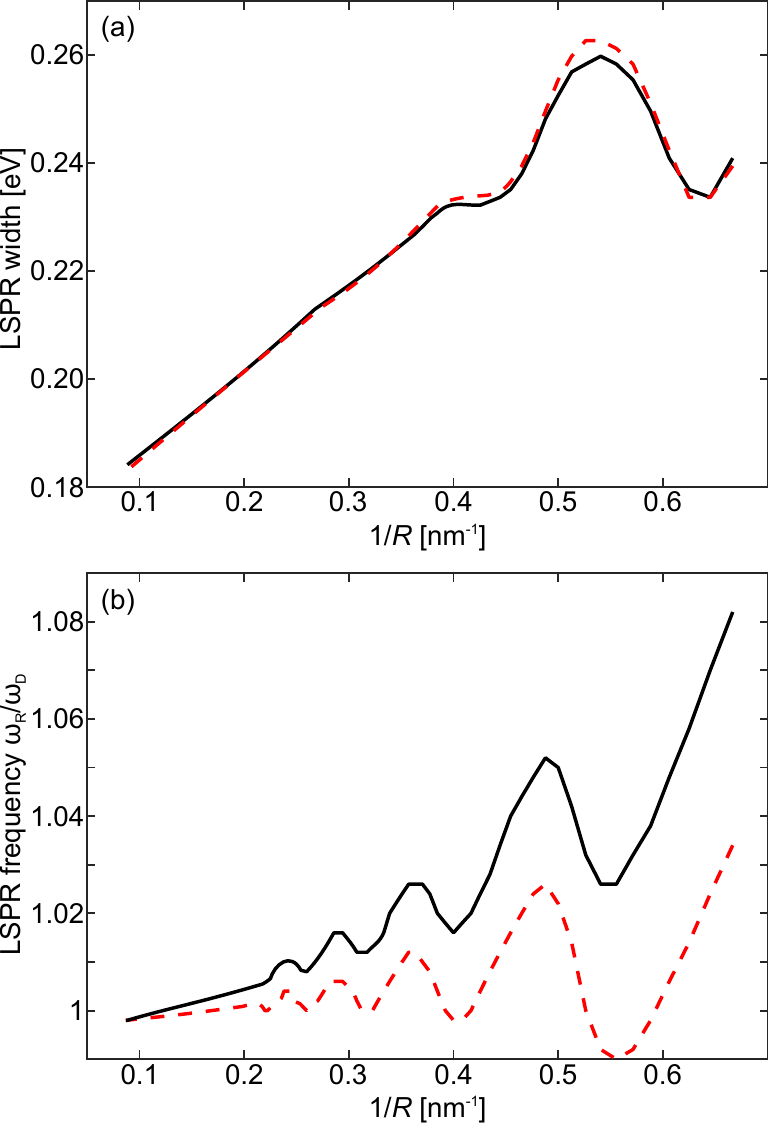}
\caption{(a) Width and (b) Energy of the LSPRs of low-electron-density Cu$_{2-x}$S as a function of $R^{-1}$ for $\beta_T=\beta_{T}^{(r)}$ and two values of $R_0$, $R_0=0$ (dashed line) and $R_0=R_{0}^{(r)}$ (continuous line). The plasmon width has a linear scaling with $R^{-1}$ up to $R^{-1} \approx 0.5$
nm$^{-1}$, and, correspondingly, the plasmon energy only increases slightly in this case. Due to the small value of $\tilde R_0$, QSE are small in the nanometer range of radii.}
\label{fig-CuS}
\end{figure}

We finally analyze the case of Cu$_{2-x}$S QDs doped to hole densities on the order of $n_h \simeq 1 \times 10^{21}$ cm$^{-3}$ dispersed in tetrachloroethylene (TCE), one of the systems where LSPRs arising in low carrier density nanocrystals were earlier reported \cite{NMat_10_361_luther}. The values of the parameters are $\epsilon_{\infty}=1$, $\epsilon_m=2.28$, $m^{*}=0.8$. We chose $\gamma_b=$0.17 eV in order to approximately reproduce the width of the $R=3$ nm crystal of \cite{NMat_10_361_luther}. 
Using $n_h=1 \times 10^{21}$ cm$^{-3}$, we obtain $r_s=11.7$, $v_F=0.45 \times 10^{6}$ m s$^{-1}$,  $\omega_p=1.31$~eV, $R_0^{(r)}=0.23$~nm, $\tilde R_0^{(r)}=0.54$~nm, and $d_\mathrm{dif}^{(r)}=1.33$~nm. 
Notice that $\tilde R_0^{(r)}$ is smaller by a factor of 0.5 and  $d_\mathrm{dif}^{(r)}$ is smaller by a factor of 0.3 than the corresponding values 
for our previous system, both systems having equal carrier densities. Consequently, we expect smaller effects of QSE and diffuse surface scattering for
 Cu$_{2-x}$S in TCE than for low-electron density ZnO in toluene.
Figure~\ref{fig-CuS}a,b shows, respectively, the width and the energy of the LSPRs for $R$ in the nanometer range and $\beta_T=\beta_{T}^{(r)}$, neglecting QSEs ($R_0=0$), and including the QSEs with $R_0=R_{0}^{(r)}$. 
What happens in this system is that the large value of the bulk damping $\gamma_b$, a factor of 1.7 larger than for photodoped ZnO, with nearly equal Fermi velocities, makes the damping length of the oscillations to be short. We therefore find a large range of values of $R$ where the width scales linearly with $R^{-1}$ and, correspondingly, a very small energy shift is found at these radii. Compared to our previous system, we notice that
the width of the resonances increases with respect to $\gamma_b$ by
a factor of 1.5 at most while factors of 3 were found in Figure~\ref{fig-width-n10} for $\beta_T=\beta_{T}^{(r)}$.
Moreover, as seen in Figure~\ref{fig-CuS}b, the small value of $R_0^{(r)}$ in this system also makes QSEs to be small and, consequently, the energy of the LSPRs increases by 8$\%$ while it is 30$\%$ for the case of low-electron density ZnO with the same ratio of $R_0/R_{0}^{(r)}$ and $\beta_T/\beta_{T}^{(r)}$, shown in Figure~\ref{fig-wdiff-n10}b.

%%%%%%%%%%%%%%%%%%%%%%%%%%%%%%%%%%%%%%
\section{Conclusions}
%%%%%%%%%%%%%%%%%%%%%%%%%%%%%%%%%%%%%%

In this article, we have investigated the role that effects of quantization due to size and diffuse surface scattering play in modifying the LSPRs of low-density carrier nanospheres as a function of size, using the theory expounded in detail in Ref. \cite{JPCL_6_1847_carmina}. The two key parameters of the theory are the length $R_0$ giving the strength of the QSEs and the velocity $\beta_T$ of the electronic excitations, entering in the length scale for diffuse surface scattering $d_\mathrm{dif}$.
While in our theory the QSE itself only produces a blueshift in energy of the LSPRs with particle size, the diffuse surface scattering mechanism gives to both energy and linewidth an oscillatory-damped behavior with characteristics lengths that depend on the carrier density and on other material parameters as well. Thus, the evolution of the LSPRs with particle size at the nanometer scale is very dependent on the relation of size to these lengths, which we illustrated with several examples. 
Our calculations for the energy shift of the resonances (relative to the Drude value) showed a critical dependence on these lengths as we obtained 80$\%$ blueshift for ultralow electron density ZnO spheres in toluene and 8$\%$ blueshift for Cu$_{2-x}$S spheres of the same radii in TCE. Significant differences in the magnitude of the plasmon width (relative to the bulk value) are also found among the investigated systems.
The variety of behaviors we found for the LSPRs could be useful for designing plasmonic devices based on doped semiconductor nano structures having desired properties.

\begin{acknowledgement}
RCM acknowledges financial support from the Spanish Mineco via the project MAT2014-53432-C5-5-R. 
TJA thanks the Foundation of Polish Science for support via the project HOMING PLUS/2013-7/1, the Polish Ministry of Science and Higher Education for support via the Iuventus Plus project IP2014 000473, and the Polish National Science Center via the project 2012/07/D/ST3/02152. 
TJA and SPA acknowledge financial support from the Swedish Foundation for Strategic Research via the 
Functional Electromagnetic Metamaterials for Optical Sensing project SSF~RMA~11.
\end{acknowledgement}

%\begin{suppinfo}
% If needed a description will go here.
%\end{suppinfo}

\providecommand{\latin}[1]{#1}
\providecommand*\mcitethebibliography{\thebibliography}
\csname @ifundefined\endcsname{endmcitethebibliography}
  {\let\endmcitethebibliography\endthebibliography}{}

\newpage

%\begin{tocentry}
%\end{tocentry}

\noindent
{\Large \textbf{For Table of Contents Use Only}}

\noindent
{\Large Diffuse Surface Scattering and Quantum Size Effects in the Surface Plasmon Resonances of Low Carrier Density Nanocrystals}

\noindent
R. Carmina Monreal, Tomasz J. Antosiewicz, and S. Peter Apell

\begin{figure}
\includegraphics{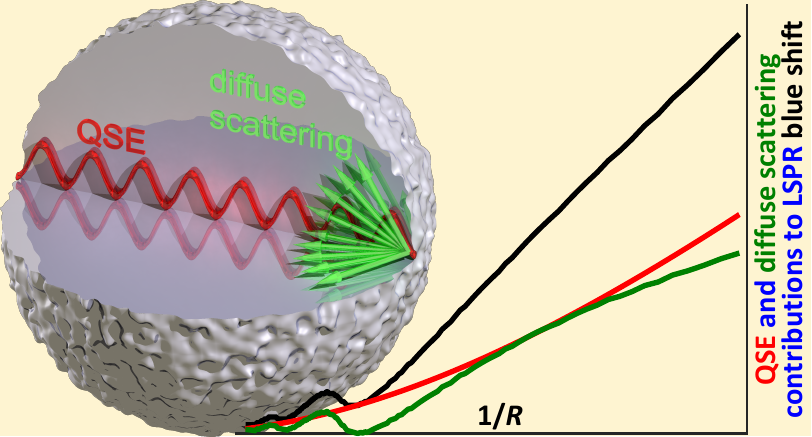}
\end{figure}

\end{document}